\documentclass[structabstract]{aa}  
\usepackage{graphicx}
\usepackage{txfonts}
\usepackage{natbib}

\usepackage{color}

\newcommand{\gr}{$\gamma$-ray}

\begin{document}
   \title{Galactic sources of $E>100$~GeV gamma-rays seen by {\it Fermi} telescope}

    \author{A. Neronov
          \inst{1}
          \and
          D.Semikoz\inst{2,3}
          }

   \institute{ISDC Data Centre for Astrophysics, Ch. d'Ecogia 16, 1290, Versoix, Switzerland \\
              \email{Andrii.Neronov@unige.ch}
         \and
             APC, 10 rue Alice Domon et Leonie Duquet, F-75205 Paris Cedex 13, France \and
Institute for Nuclear Research RAS, 60th October Anniversary prosp. 7a, Moscow, 117312, Russia
\\
             \email{Dmitri Semikoz <dmitri.semikoz@apc.univ-paris7.fr>}
             }


 
  \abstract
   {}
{We perform a search for sources of  $\gamma$-rays with energies $E>100$~GeV at low Galactic latitudes $|b|<10^\circ$  using the data of \textit{Fermi} telescope.}
{To separate compact $\gamma$-ray sources from the diffuse emission from the Galaxy, we use the  Minimal Spanning Tree method with threshold   of $\ge 5$ events in inner Galaxy (Galactic longitude $|l|<60^\circ$) and  of $\ge 3$ events in outer Galaxy. Using this method, we identify 22 clusters of very-high-energy (VHE) \gr s, which we consider as "source candidates". 3 out of 22 event clusters are expected to be produced in result of random coincidences of arrival directions of diffuse background photons.   To distinguish clusters of VHE events produced by real sources from the background we perform likelihood analysis on each source candidate}
{We present a list of 19 higher significance sources for which the likelihood analysis in the energy band $E\ge 100$~GeV  gives Test Statistics (TS) values above $25$. Only 10 out of the 19 high-significance sources can be readily identified with previously known VHE \gr\ sources. 4 sources could be parts of extended emission from known VHE \gr\ sources. Five sources are new detections in the VHE band. Among these new detections we tentatively identify one  source as a possible extragalactic source PMN J1603-4904 (a blazar candidate), one as a pulsar wind nebula around PSR J1828-1007. High significance cluster of VHE events is also found at the position of a source coincident with the Eta Carinae nebula. In the Galactic Center region, strong VHE \gr\ signal is detected from Sgr C molecular cloud, but not from the Galactic Center itself. }
{}

\maketitle

\section{Introduction}

Systematic exploration of the Galaxy in the very-high-energy \gr\ band (photon energies above 100 GeV) has started with HESS survey of the inner  Galactic plane \citep{HESS_survey}.  Some $\simeq 60$ Galactic VHE \gr\ sources of different types, such as supernova remnants (SNR), pulsar wind nebulae (PWN), \gr\ loud binary systems (GRLB), molecular clouds (MC), as well as unidentified sources without obvious low energy counterparts are known today\footnote{See e.g. http://tevcat.uchicago.edu/, http://www.mppmu.mpg.de/~rwagner/sources/ for the VHE source catalogues.}.  Survey of the Northern sky with MILAGRO \citep{MILAGRO_survey} has revealed several brightest VHE \gr\ sources in the Cygnus region as well as an unidentified source MGRO J1908+06 which is most probably a pulsar wind nebula associated to associated to a \gr\ pulsar PSR J1907+0602 \citep{mgro1908}. Outside the inner $\sim 60^\circ$ of the Galactic Plane as well as in the Galactic Bulge, no survey comparable by depth to the HESS survey is available so far, so that significant fraction of the Galaxy remains unexplored in the VHE \gr\ band. 

The main difficulty for surveying large parts of the sky in the VHE \gr\ band is the narrow field of view  (3-5 degrees) of the ground-based imaging Cherenkov telescopes, which are the most sensitive instruments in this energy band. In this respect, Large Area Telescope (LAT) on board of {\it Fermi} satellite is complementary to the ground based \gr\ telescopes. It has much smaller collection area ($\sim 1$~m$^2$ above 100 GeV) than the ground based telescopes, but its field of view is $\simeq 2.5$~sr and it surveys the entire \gr\ sky every 3.2 hr\footnote{http://www-glast.slac.stanford.edu/\\
software/IS/glast\_lat\_performance.htm}. Typical flux of VHE \gr\ sources is below $\sim 10^{-11}$~erg/cm$^2$s, so that at the energies above 100~GeV no more than $\sim 10$ photons could be detected from the brightest sources. However, the instrumental background of LAT in the VHE band is extremely low, so that a signal with several photon statistic can be significantly detected. 
\citet{neronov10,neronov_IC310} used this fact to produce a survey of extragalactic sky (Galactic latitudes $|b|>10^\circ$ at the energies $E>100$~GeV). 

 Two different methods analysis of the low statistics signal in the energy range above 100~GeV  were applied to {\it Fermi} data at high Galactic latitudes \citep{neronov10,neronov_IC310}. First, individual \gr\ sources could be detected with significance close to 5$\sigma$ already for  $N\ge 3$ \gr s from the source direction, via auto-correlation analysis of photon arrival directions \citep{neronov_IC310}. Apart form several brightest blazars, only one previously unknown VHE \gr\ source, a radio galaxy IC 310, has been discovered by LAT above 100~GeV in this way.  Detection of this source is now confirmed by observations with the ground-based \gr\ telescope MAGIC \citep{IC310_MAGIC}. Otherwise, cross-correlation analysis of arrival directions of $E>100$~GeV photons with a pre-defined source catalog could be used to select "source candidates" in such a way that most of the sources (e.g. $\ge 90$\%) correlating with the $E>100$~GeV \gr\ arrival directions would be real  sources with only a minor possibility of false detections \citep{neronov10}.

In what follows we perform a search for Galactic VHE \gr\ sources at $|b|\le 10^\circ$ using {\it Fermi}/LAT data in the VHE range $E>100$~GeV, using the auto-correlation analysis method. The main difference between the analysis at high and low Galactic latitudes is in the level of diffuse background, especially in the inner Galaxy. The increase of the level of diffuse background leads to the increase of the threshold for the source detection. This complicates identification of isolated clusters of \gr\ events which might be produced by well-defined \gr\ sources, rather then by inhomogeneities of the diffuse emission. We use the Minimal Spanning Tree (MST) method (see \citet{Canipana:2008zz,Massaro:2009ib} for application of this method for the analysis of \gr\ data) to identify compact clusters of VHE events.  To minimize the probability of mis-identification of excesses in the diffuse emission as isolated \gr\ sources, we consider only clusters of $\ge 3$ events in the outer Galaxy (Galactic longitude $|l|\ge 60^\circ$) and  clusters of $\ge 5$ events in the inner Galaxy. 

We present a catalog of 19 sources at low Galactic latitude $|b|<10^\circ$, found using MST method  above 100 GeV in {\it Fermi}/LAT data. Only 10 out of the 19 sources could be readily identified with  VHE \gr\ emitters detected by HESS, MAGIC, Veritas  and MILAGRO telescopes. We discuss the nature of the remaining 9  sources and find that four of them might be associated to the known extended VHE sources, while the other 5 are new detections in the VHE band.

\section{Data selection and data analysis}

In our analysis we use the data taken in the period between August 4, 2008 and September 2, 2010 by the Large Area Telescope (LAT) on-board of {\it Fermi} satellite. We filter the data using {\it gtselect} and {\it gtmktime} tools which are part of the standard data analysis package provided by {\it Fermi} collaboration\footnote{See http://fermi.gsfc.nasa.gov/ssc/data/analysis/scitools}. We retain only \gr -like events (event class 3) in the energy range above 100 GeV for our analysis. This leaves 6995 events distributed over the whole sky.  In our MST analysis we concentrate on the subset of  2497 events at Galactic latitude $|b|<10^\circ$. In the likelihood analysis we restrict the energy range to 100-400 GeV, disregarding the higher energy events, for which LAT instrument characteristics are not included in the standard LAT data analysis package. LAT point spread funciton (PSF) at the energies above 100 GeV has the radius $\sim 0.1^\circ$. At the same time, typical distance between the bright sources visible with LAT in the VHE band is $\ge 1^\circ$. This means that, contrary to the lower energy band data analysis, overlap of the PSF between different sources normally does not affect the estimates of source fluxes in the standard likelihood analysis. At the same time, statistics of the signal from both point sources and from the diffuse Galactic and extragalactic backgrounds is rather low in the VHE band. The main uncertainty for the estimate of the fluxes of VHE sources is introduced by the uncertainty of the measurement of the diffuse background levels. To have a good estimate of the diffuse backgrounds at the position of each source we choose a sufficiently large region of the radius $10^\circ$ around the source of interest, so that the background signal statistics is at least at the level of $\sim 10^2$ photons. In the likelihood analysis we take into account all the sources at the positions of clusters of  VHE photons listed in Table \ref{tab:ts25}.

\section{Search for clusters   in the {\it Fermi} data at  $E\ge 100$~GeV with Minimal Spanning Tree method}
\label{sec:tree}

In order to find all possible point-like and/or extended isolated sources of VHE  $\gamma$-rays in the Galaxy, we used
the MST method~\citep{Canipana:2008zz,Massaro:2009ib}. In this way we identify all compact clusters of VHE events, which we consider as "source candidates" for the subsequent analysis. Event clusters identified in the MST method are characterized by maximal angular distance $\theta$ between any two events forming the "branch" of the "tree"  and the number $n$ of events in the cluster ("tree"). 

As it is mentioned above, there are $2497$ VHE photons in the Galactic Plane with $|b|<10^\circ$. Imposing a restriction that $\theta$ should be smaller than a  pre-defined boundary value $\theta_0$, one finds that most of the VHE \gr\ events would belong to singlets and that individual event clusters are well separated from each other. For example, choosing  $\theta_0=0.4^\circ$, we find that only 786 events are in the clusters with $n\ge 2$. 

Isolated clusters of events in a given direction on the sky could be formed either because of the presence of a relatively bright \gr\ source at the position of the cluster or due to random coincidence of arrival directions of photons from diffuse (Galactic and extragalactic) sky background. 
To estimate the number of clusters formed clusters due to random coincidences of arrival directions of photons from the diffuse background, we generate 1000 Monte-Carlo (MC) simulated realizations of diffuse \gr\ emission and count the number of event clusters in the MC data sets.   The diffuse Galactic plus extragalactic background is modeled in the following way.  First, we remove a cluster of events around the position of a known bright source, Crab, from the real data set (29 events). Next, each MC realization is produced from  the remaining set of real \gr\ events with energies above 100~GeV,  detected by LAT. We leave Galactic latitude $b$ the same  for all events, but re-distribute the Galactic longitude $l$ within $\pm 10^\circ$ from real arrival direction of the photon. This procedure erases all the real event clusters in the data, which might be due to the real (point-like) \gr\ sources. At the same time, it does not change the overall pattern of diffuse emission from the Galaxy, which typically changes on the angular scale larger than $\sim 10^\circ$ in $l$ direction. Leaving $b$ unchanged we preserve the Galactic diffuse background pattern, which is strongly concentrated around the Galactic Plane, in the latitude range $|b|\lesssim 1^\circ$. Clusters of events present in the MC data sets generated in such a way could be only due to the random coincidences of photon arrival directions. Note that photons from the real Galactic \gr\ sources are re-distributed and contribute to the diffuse emission in the MC data sets. However, the number of photons which initially belonged to the real clusters of events is small compared to the total number of events ($\sim 10\%$), so that the presence of real source photons leads to just a slight over-estimate of diffuse background flux in the MC data sets.

We compare the estimate of the number of clusters of VHE \gr\ events with given multiplicity $n_0$ and within angular distance $\theta\le \theta_0$ in the MC data sets with the number of clusters with the same $n_0, \theta_0$ in the real data.  Taking into account significant difference between the levels of diffuse background in the inner and outer Galaxy ($|l|<60^\circ$ and  $|l|>60^\circ$, respectively),  we use different criteria for selection of clusters which might be due to the real sources in the outer and inner Galaxy.

\begin{figure}[htbp]
\begin{center}
\includegraphics[width=0.7\linewidth,angle=-90]{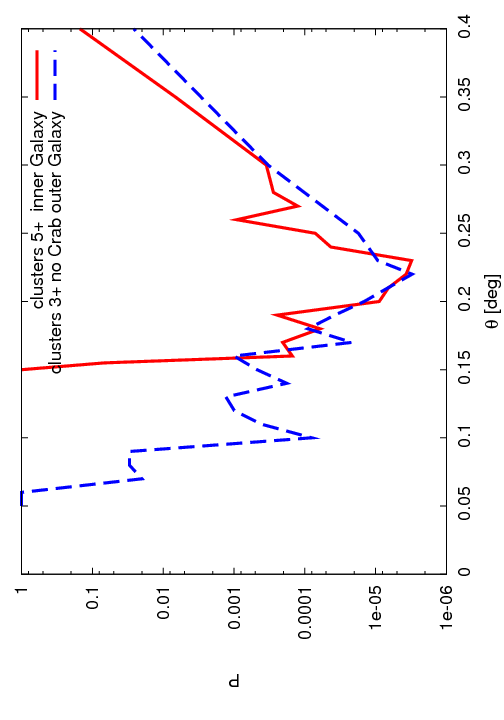}
\end{center}
\caption{Probability to find the observed number of clusters of VHE events as a function of angular separation of events in the cluster $\theta$.  Solid red curve is for  $n_0\ge 5$ clusters in the inner Galaxy $|l| \le 60^\circ$.
 Dashed blue curve is for   $n_0\ge 3$ clusters in the outer Galaxy $|l| \ge 60^\circ$.}
\label{fig:probability}
\end{figure}

For outer Galaxy we find that clusters with two events only are consistent with background at any $\theta_0$. At the same time, the number of clusters with $n_0\ge 3$ in the real data is much larger than that in the MC data sets.  
This implies that significant amount of $n_0\ge 3$ event clusters is formed by \gr s from real compact sources of \gr s.  The probability that clusters of $n\ge 3$ events found in the outer Galaxy are produced in result of random coincidence of background photon arrival directions is shown as a function of $\theta$ in Fig. \ref{fig:probability} by the blue dashed line.  The probability reaches minimum $P=3 \cdot 10^{-6}$ at the angle $\theta=0.22^\circ$. 11 clusters with this $\theta$ are found in the real LAT data (including Crab), while 1.45 expected to be formed by chance, on average. 

In the inner Galaxy, the number of event clusters with $n=3$ and $n=4$ is comparable to the one expected from MC simulations. At the same time, the number of event clusters with $n\ge 5$ in the real data significantly exceeds the one found in the MC simulaitons (see Fig. \ref{fig:probability}).  The chance probability to find the observed number of clusters has minimum $P=3.1 \cdot 10^{-6}$ at $\theta_0=0.23^\circ$ with 12 clusters in data vs. 1.8 expected by chance. 

In total we have 22 source candidate in the inner and outer Galactic Plane  and, on average,  3.25 of them expected to be by chance.


To separate real sources from possible false detections, we perform the standard unbinned {\it Fermi}/LAT likelihood analysis for each source candidate. In real sources the VHE events in the cluster are distributed following the PSF, while in the background clusters photons are less likely to follow the PSF. We assume that the most likely false detections are source candidates with the smallest values of the Test Statistics (TS) values \citep{mattox}. Comparing distribution of the TS values of the source candidates with the results of MC simulations we find that correct proportion of the false detections to the total number of clusters if found when $TS\simeq 25$ is chosen to separate real sources from the possible false detections. Same value was chosen the 1-st year Fermi catalog  \citep{fermi_catalog} to distinguish real sources from the fluctuations of background.

\section{LAT VHE source catalog}

\begin{table*}
\begin{tabular}{|l|l|l|l|l|l|l|l|l|}
\hline
&VHE&RA &DEC& TS& $F^\dagger$ &$n$ &Name& Type\\
\hline
1  & J0521+2114 &80.37817      &21.19834    &36.3   &$7.9\pm 4.2$   &3  &  {\bf VER J0521+211}  & AGN \\
2  & J0534+2200 &83.64470      &22.01954    &385.0 &$57\pm 11$	    &29&  {\bf  Crab}                      &PWN   \\
3  & J0616+2246 &94.27895      &22.70302    &32.7   &$9.4\pm 4.7$   &3  &  {\bf IC 443}                    &SNR   \\ 
4  & J1025-5808  &156.31574    & -58.136402  &27.9   &$8.6\pm 4.7$   &3  &      &   \\ 
5  & J1044-5949  &161.2390      &-59.72588   &25.5   &$8.4\pm 4.6$   &3  &  Eta Carinae(?)              & GRLB (?)	       \\
6  & J1603-4909  &240.91667    &-49.16073   &37.9   &$9.5\pm 4.5$  &5  &  PMN J1603-4904         & AGN\\	 
7& J1634-4737  &248.58186    &-47.619664 & 47.9 &$14.6\pm 6.2$&5  &&\\
8& J1647-4638  &251.93635    &-46.639038 & 30.5 &$12.1\pm 6.0$&5 &   &PWN\\
9& J1714-3922  &258.69515    &-39.371822  &32.7 &$13.5\pm 5.9$&5+2  &  {\bf RX J1713.7-3946} & SNR\\
10& J1744-2921  & 266.1650     &-29.36567    &60.7  &$20.1\pm 7.6$&5  & Sgr C                       & MC(?)\\
11&  J1828-0959 &277.1416      &-10.03979    &53.7  &$16.1\pm 7.1$&5   & PSR J1828-1007(?)     &PWN(?)\\
12& J1837-0659  &279.41075    &-6.9947278  &28.0  &$13.3\pm 6.5$&3+8  &  {\bf HESS 1837.5-069 }&PWN	  \\
13& J1838-0646  &279.69634    &-6.7674015  & 36.2 &$14.0\pm 6.3$&5  &{\bf AX J1838.0-0655}    &PWN\\
14& J1839-0550  &279.8174      &-5.842005     &25.3  &$11.9\pm 5.8$&7  & 1FGL J1839.1-0543     &  \\
15& J1857+0252 &284.25073    &2.8717481    &40.9  &$20.3\pm 7.4$&7 & {\bf PSR J1856+0245}   &PWN\\
16& J2001+4351 &300.2823      &43.84117     &38.9    &$5.3\pm 3.2$  &3 &  {\bf VCS1 J2001+4352}&AGN\\  
17& J2019+4048 & 304.82017   &40.812433   &28.1   &$9.0\pm 4.9$  &3+1 &{\bf VER J2019+407}      &SNR/PWN\\
18& J2320+5911 &350.17173      &59.19705    &37.5    &$6.2 \pm 3.6$ &3 &                                            &\\
19& J2347+5142 &356.7813      &51.68506      &37.7   &$5.5\pm 3.2$  &4 &  {\bf 1ES 2344+514 }      &AGN  \\
\hline
\end{tabular}
\caption{Low Galactic latitude LAT sources above 100~GeV. $F$ is flux normalization at $100$~GeV in units of $10^{-16}$~cm$^{-2}$s$^{-1}$MeV$^{-1}$. $n$ in the multiplicity of events in the cluster associated to the source. }
\label{tab:ts25}
\end{table*}         

The list of 19 source candidates  with $TS\ge 25$ found at the places of clusters of  VHE events is given in Table \ref{tab:ts25}.  Three event clusters with $TS\le 25$ are listed in Table \ref{tab:false}. We consider these three low-significance source candidates as the most likely false detections and exclude them from the high-confidence VHE source list given in Table \ref{tab:ts25}. 

\begin{table}
\begin{tabular}{|l|l|l|l|l|l|}
\hline
&RA &DEC& TS& $F^\dagger$ &$n$\\
\hline
1  &157.67473   &  -58.17904 & 17.1  &$2.3\pm 2.2$    & 3\\
2  &244.28805    &-51.043775 &23.0  &$6.3\pm 4.2$  &5    \\
3 &310.23197    &42.573738  &22.0     &$5.6\pm 3.3$ & 3   \\
\hline
\end{tabular}
\caption{Event clusters with $TS\le 25$. Flux $F$ and multiplicity $n$ columns are the same as in the Table  \ref{tab:ts25}.}
\label{tab:false}
\end{table}         

From Table \ref{tab:ts25} one can see that significant fraction (10 out of 19) of the high TS value sources are positionally coincident with the known VHE \gr\ sources, marked in bold\footnote{see http://tevcat.uchicago.edu/ for a VHE source catalog.}.  The remaining sources could not be directly identified as the counterparts of known VHE \gr\ sources. 

This might be due to several possible reasons. First, for the sources outside the inner $\pm 60^\circ$ of the Galactic Plane, no systematic survey of the depth comparable to the HESS Galactic Plane survey \citep{HESS_survey} is available. It is possible that the sky region around the source was just never observed with the ground-based \gr\ telescopes. Second, for the sources in the inner Galaxy, at Galactic latitude $|b|\gtrsim 2^\circ$ the depth of the HESS Galactic Plane survey decreases due to the decrease of the effective collection area at the periphery of the HESS field of view. Spectra of some of the sources detected by LAT at the energy above 100~GeV might have a high-energy cut-off just above 100~GeV, so that they escape detection by the ground based \gr\ telescopes with energy thresholds much above 100~GeV. Finally, many VHE \gr\ sources detected by the ground-based \gr\ telescopes are extended and have energy dependent morphology. It is possible that centroid of the source shifts with energy so that position of the source found at TeV energies does not coincide with that found in the 100~GeV energy band. 

Below we discuss the nature of the 9 new VHE \gr\ sources detected by LAT in more details.

\section{Remarks on individual sources}

\begin{figure*}
\includegraphics[width=\linewidth]{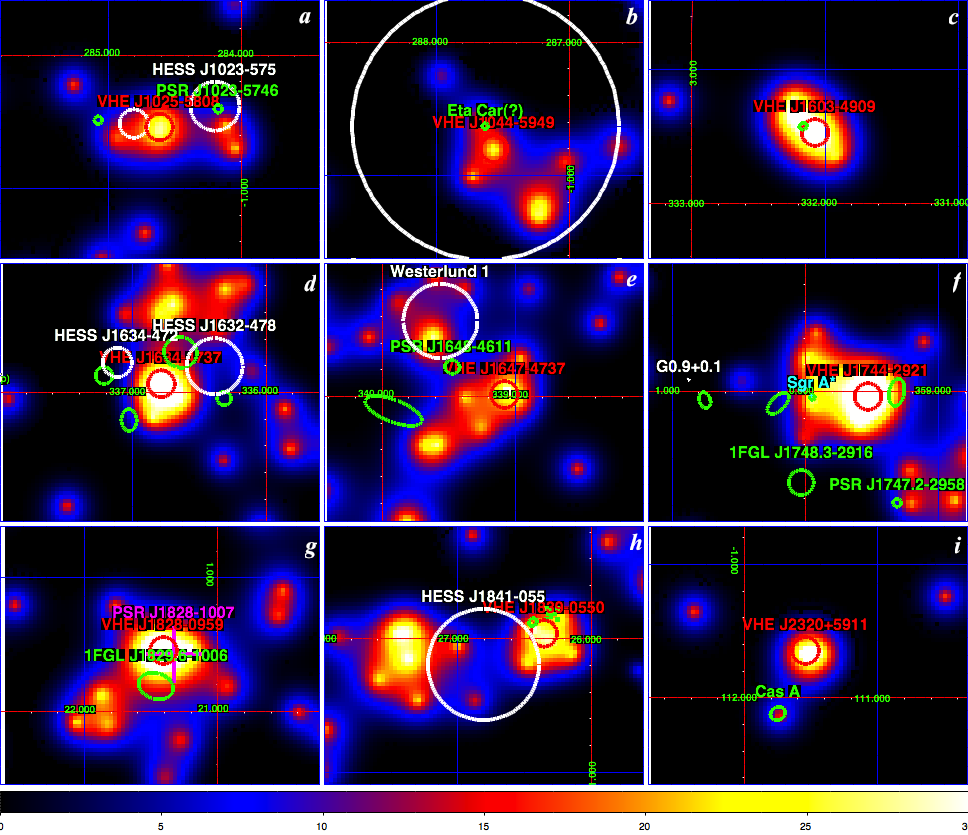}
\caption{TS maps in the 100-400 GeV band around positions of new VHE \gr\ sources from Tables \ref{tab:ts25}. Green ellipses show the 95\% uncertainties of positions of sources from the 1-st year Fermi catalog. Cyan circles show position uncertainties of the sources from HESS Galactic Plane survey. Red crosses mark the positions of VHE source centroids in 100-400~GeV band in the cases when $TS\ge 30$. }
\label{fig:highTS}
\end{figure*}

Fig. \ref{fig:highTS} shows the maps of TS values in the 100-400~GeV energy band around the positions of VHE event clusters for which no direct identification with a known VHE \gr\ source is possible. In the same figure we show the uncertainties of positions  of the sources from the 1-st year {\it Fermi} catalog \citep{fermi_catalog} and from the HESS survey of the inner Galactic Plane \citep{HESS_survey}. One can see that in four cases, of VHE J1025-5808, VHE J1634-4737, VHE J1744-2921 and VHE J1839-0550 known extended VHE \gr\ sources are present in the direct vicinity of {\it Fermi} VHE \gr\ source. This implies that the source found by LAT in the 100-400~GeV energy band might be a part of the degree-scale extended VHE \gr\ emission region. The discrepancy between the positions of {\it Fermi} and HESS sources might be due to the energy dependence of the source morphology. At the same time, five new sources  are not directly adjacent to the known extended VHE \gr\ sources, so that they should be considered as new isolated VHE \gr\ sources.

\vskip0.3cm
\textit{VHE J1025-5808} is situated between two \gr\ pulsars, PSR J1028-5819 (shown by the green unlabeled circle in Fig. \ref{fig:highTS}a) and PSR J1023-5746, see Fig. \ref{fig:highTS}a.  PSR J1023-5746 is in the center of an extended TeV band source at the position of Westerlund 2 stellar cluster \citep{westerlund2}. VHE J1025-5808 is displaced from the position of the HESS source toward PSR J1028-5819, so that it is not clear if the new VHE source is a part of the extended emission from the Westerlund 2 cluster region or it is an independent source. HESS collaboration has recently reported the detection of a new source most probably associated to the PSR J1028-5819 \citep{VHE1025}, which is almost coincident with VHE J1025-5808, but is further displaced toward the position of the pulsar PSR J1028-5819. Unlabeled white circle in Fig. \ref{fig:highTS}a marks the position of the newly discovered HESS source.

\vskip0.3cm
\textit{VHE J1044-5949} is a VHE event cluster associated to the {\it Fermi} source 1FGL J1045.2-5942. This source is positionally coincident with the Eta Car colliding wind gamma-ray loud binary (GRLB) \citep{agile_etacar}. Absence of the variability of the {\it Fermi} source on the orbital time scale of Eta Car  did not allow its firm identification with the Eta Car system \citep{fermi_etacar}. From panel {\it b} of Fig. \ref{fig:highTS} one can see that  the centroid of the VHE source appears to be displaced in the south-west direction from the binary system itself, shown as a green circle. This might indicate that the VHE emission is extended and is, in fact produced in the nebula powered by Eta Car, rather than at Eta Car itself. The extent of the Eta Car nebula is shown by a white circle in Fig. \ref{fig:highTS}b. \gr\ emission from the nebula would also explain the absence of variability of the {\it Fermi} source on the Eta Car orbital time scale.

\vskip0.3cm
\textit{VHE J1603-4909} is identified with 1FGL J1603.8-4903, which, in turn was identified as an AGN by  \citep{Fermi_AGN} based on the correlation with a radio source PMN J1603-4904. TS map shown in Fig. \ref{fig:highTS}c shows a clear point-like source detected above 100~GeV at the position of the known {\it Fermi} source.  Thus, the new VHE source is, most probably an AGN behind the Galactic Plane. 

\vskip0.3cm
{\it VHE J1634-4737} is situated in the middle of a cluster of unidentified {\it Fermi} and HESS sources, 1FGL J1636.4-4737c, 1FGL J1632-4802, 1FGL J1632.6-4733c, 1FGL J1635.7-4715c, HESS J1632-472, HESS J1632-478, see Fig. \ref{fig:highTS}d. All these sources are unidentified. Presence of a large number of \gr\ sources with positions shifting with the change of energy points to the common origin of both lower energy ($\sim 1$~GeV) sources from the 1-st year {\it Fermi} catalog, of the intermediate energy (100~GeV) source VHE J1634-4737 and of the high-energy ($\sim 1$~TeV) HESS sources. It is possible that all these sources form an extended emission region with energy  dependent morphology. The source power of the extended emission remains, however, unknown.

\vskip0.3cm
{\it VHE J1647-4737} is most probably an extended source at the position of Westerlund 1 massive star cluster. Extended VHE \gr\ emission from this region was recently reported by HESS \citep{ohm09}. 1-st year {\it Fermi} catalog lists a source 1FGL J1648.4-4609 which is 0.4 degrees from the position of VHE J1647-4634, see Fig. \ref{fig:highTS}e. The {\it Fermi} catalog source is positionally coincident with the pulsar PSR J1648-4611. The extended emission around VHE J1647-4634 might, therefore, be the PWN around PSR J1648-4611.

\vskip0.3cm
{\it VHE J1744-2921} is close to the Galactic Center. An unidentified Fermi source 1FGL J1744.0-2931 is found close to the position of the VHE source, see Fig. \ref{fig:highTS}f.  Centroid of the VHE emission is positionally coincident with the core of Sgr C molecular cloud, which exhibits bright X-ray reflection features \citep{nakajima09}. It remains to be seen why emission from Srg C cloud strongly dominates the Galactic Center emission at 100~GeV energy, while giving sub-dominant contribution to the emission from the Galactic Ridge at slightly lower and slightly higher energies. 

\vskip0.3cm
{\it VHE J1828-0958} is close to 1FGL J1829.6-1006, but the distance between the centroid of the VHE source and the position of the 1FGL source is slightly larger than $0.25^\circ$, see Fig. \ref{fig:highTS}g. A pulsar PSR J1828-1007 at the distance 0.1 degree from the source is a probable low-energy counterpart of the source.  Shift in the source position from low to high energies and presence of a pulsar favor the interpretation of the source as a PWN. 

\vskip0.3cm
\textit{VHE J1839-0550} is an event cluster directly adjacent, but not coincident with a very extended HESS source  HESS J1841-055, see Fig. \ref{fig:highTS}h. It is interesting to note that a similarly high TS event cluster is found on an opposite side of HESS J1841-055. However, this event cluster is not listed in Table \ref{tab:ts25} as a high-significance VHE \gr\ source because it is more extended, so that the distance between photons is larger than $\theta\le 0.23$ adopted in our MST method. This demonstrates an important fact that the list of VHE \gr\ sources presented in Table \ref{tab:ts25} is not the exhaustive list of VHE \gr\ sources detected by LAT. It is only the list of VHE sources which can be identified using MST method which we adopted for selection of VHE source candidates.  

\vskip0.3cm
{\it VHE J2320+5910 } is a source  $0.5^\circ$ from Cas A SNR, see Fig. \ref{fig:highTS}i.  Presence of a high-TS excess above 100~GeV close to Cas A appears puzzling, taking into account the fact that the region around Cas A was extensively observed by the ground based \gr\ telescopes with energy threshold of just several 100~GeV.  It is possible that the source escaped detection because it was always at large off-axis angles for the ground-based \gr\ telescopes MAGIC and Veritas, which have rather narrow field of view ($\simeq 1.5^\circ$ in radius). Otherwise, the spectrum of the source might have a cut-off at the energies around $\sim 100$~GeV so that it is not detectable for the instruments with energy thresholds of several 100 GeV.  Re-observation of the Cas A region with MAGIC-II telescopes \citep{IC310_MAGIC} which have low-energy threshold below 100~GeV might clarify is the observed high-TS excess is a real source or it is an event cluster formed due to random coincidence of diffuse photon arrival directions.

\section{Discussion and Conclusions}

We have performed a search for the sources of VHE gamma-rays  with $E>100$ GeV  at  low Galactic latitudes $|b|<10^\circ$. Using two years of LAT Fermi data, we applied the Minimal Spanning Tree method to identify clusters of VHE \gr\ events superimposed on the diffuse emission from the Galaxy. Higher level of diffuse Galactic background in the inner Galaxy leads to a higher occurrence rate of event clusters produced by the random coincidences of arrival directions of the background photons. Taking this into account, we have split the Galactic Plane on the inner and outer regions and analyzed separately the two regions. To separate event clusters produced by real sources from the clusters of background events, we imposed a requirement that minimal number of events in the clusters should be $n\ge 5$ in the inner galaxy and $n\ge 3$ in the outer Galaxy.

Using the MST method, we have identified 22 most significant event clusters at low Galactic latitudes in the LAT data. At the same time, only 3 clusters with $n\ge 5$ ($n\ge 3$) events in the inner (outer) Galaxy are expected to be background event clusters. We have performed the likelihood analysis and rejected three low significance (TS values below 25 in the 100-400~GeV band) clusters as the most likely background event clusters. The resulting list of high-significance VHE \gr\ sources identified in the LAT data is presented in the Table \ref{tab:ts25}. 10 out of the 19 sources are readily identified with the previously known VHE \gr\ sources. Another four most probably are parts of extended emission with energy-dependent morphology at the positions of very extended (degree scale) known VHE \gr\ sources. The remaining 5 sources are new detections in the VHE \gr\ band. 

One of the new VHE \gr\ sources is possibly associated to the Eta Carinae colliding wind binary system and/or an extended emission nebula powered by Eta Car. Another new source is, most probably a blazar behind the Galactic Plane. One strong source of 100~GeV \gr s is positionally coincident with the core of Sgr C molecular cloud in the Galactic Center region. One source, VHE J1828-0959 is, most probably a pulsar wind nebula. Finally, an unidentified source is detected close to Cas A supernova remnant.

It is important to note that applying the MST method with the adopted  cuts on the multiplicity $n$ of event clusters and distance $\theta$ between the events in clusters, we miss a significant number of Galactic VHE \gr\ sources detectable in the LAT data. This is clear from the fact that a large number of clusters with $\theta>0.23^\circ$ is found at the position of known extended VHE \gr\ sources. Imposing the restriction $\theta\le 0.23^\circ$, we have minimized the number of possible false detections, while missing some real sources. This means that the VHE \gr\ source list presented in Table \ref{tab:ts25} should not be considered as an exhaustive list of sources detected with LAT above 100~GeV in the sky region $|b|\le 10^\circ$.

\section*{Acknowledgement}

The work of AN is supported by the Swiss National Science Foundation grant PP00P2\_123426.

\end{document}